\documentclass[letterpaper, paper,11pt]{AAS}		

\usepackage{bm}
\usepackage{amsmath}
\usepackage{amssymb}
\usepackage{subfigure}
\usepackage[colorlinks=true, pdfstartview=FitV, linkcolor=black, citecolor= black, urlcolor= black]{hyperref}
\usepackage{overcite}
\usepackage{footnpag}			      	
\usepackage{enumitem} 
\usepackage{float}
\usepackage{tcolorbox}

\PaperNumber{25-491}

\begin{document}

\title{COMPARISON OF FORCED AND UNFORCED RENDEZVOUS PROXIMITY OPERATIONS, AND DOCKING UNDER MODEL MISMATCH}

\author{Robert Muldrow\thanks{Undergraduate Research, MAE, University of Florida, Gainesville, Florida.},  
Channing Ludden\thanks{Graduate Research, MAE, University of Florida, Gainesville, Florida.},
\ and Christopher Petersen\thanks{Principal Investigator Research, MAE, University of Florida, Gainesville, Florida.}
}

\maketitle{}

\begin{abstract}
This paper compares the required fuel usage for forced and unforced motion of a chaser satellite engaged in Rendezvous, Proximity Operations, and Docking (RPOD) maneuvers. Improved RPOD models are vital, particularly as the space industry expands and demands for improved fuel efficiency, cost effectiveness, and mission life span increase. This paper specifically examines the Clohessy-Wiltshire (CW) Equations and the extent of model mismatch by comparing predicted trajectories from this model with a more computationally complex, higher fidelity RPOD model. This paper assesses several test cases of similar mission parameters, in each case comparing natural motion circumnavigation (NMC) with comparable forced motion circumnavigation. The Guidance, Navigation, and Control (GNC) impulse maneuvers required to maintain the supposedly zero fuel CW trajectories is representative of the extent of CW model mismatch. This paper demonstrates that unforced motions are not inherently more fuel efficient than forced motions, thus permitting extended orbital operations given the higher fuel efficiency. \end{abstract}

\begin{tcolorbox}[colback=gray!10, colframe=black, 
                  arc=1.5mm, boxrule=1.5pt, width=\textwidth]
\normalsize
This is the author’s original manuscript (pre-print) of the paper 
AAS 25-491, presented at the \textit{35th AAS/AIAA Space Flight Mechanics Meeting}, 
Kaua’i, Hawaii, January 19–23, 2025.
\end{tcolorbox}

\section{Introduction}

Rendezvous, Proximity Operations, and Docking (RPOD) maneuvers form a critical component of the continued expansion into the space sector, both for commercial entities and government space agencies.\cite{Bannon2024}  RPOD technology and models enable missions involving satellite formation flight, servicing and refueling of existing satellites, removal of orbital debris, manufacturing and construction.\cite{Petersen2024, Arney2023} State-of-the-art RPOD maneuvering places importance on natural motion circumnavigation (NMCs),\cite{Locke2023} where NMCs are stable orbits of a chaser satellite orbiting a target satellite while theoretically expending zero-fuel and requiring no impulse maneuvers to maintain that circumnavigational orbit.\cite{Spencer2015}  Dynamical models are naturally utilized to determine what orbital elements and trajectory define this stable orbit. These dynamical models are also utilized for autonomous satellite guidance, navigation, and control (GNC) when satellites are required to maintain a planned trajectory and resist perturbation forces. Given computational and power limitations for satellites, implemented dynamical models under autonomous RPOD conditions must be lower fidelity and less computationally complex. This higher computational simplicity results in model mismatch. One such computationally simpler RPOD dynamical model are the Clohessy Wiltshire (CW) Equations. This paper quantifies model mismatch under the CW Equations using several test cases comparing unforced, NMCs given by the CW model with comparable forced motions, demonstrating that forced motions are, counterintuitively, more fuel efficient than unforced,  “zero-fuel” maneuvers under certain conditions. The value of this novel comparison is evident through the increasing interest in RPOD that can be observed through several upcoming and ongoing space missions, particularly with the growth of commercial space travel companies, such as SpaceX, Boeing, and Blue Origin, often joining with governmental efforts to venture into space. In one such instance, Boeing is working towards a launch carrying NASA astronauts to the International Space Station through Boeing’s Starliner spacecraft.\cite{Boeing2024} Space agencies across the world are also increasing efforts to intercept and study asteroids, as noted through the Lucy,\cite{Lucy2023} Psyche,\cite{Psyche2024} Hera,\cite{Hera2024} and Comet Interceptor\cite{Interceptor2024} missions. The Comet Interceptor mission is of notable complexity, involving three spacecraft coordinating to observe a satellite from multiple angles.\cite{Interceptor2024} All above missions require extensive and precise course correction under highly limited fuel surpluses. The necessity of highly fuel-efficient orbital paths is apparent under these constraints.

The dynamical problem of model mismatch under the CW Equations has been examined in previous studies, however none of these studies examine this model mismatch with respect to forced and unforced motion. Research conducted by J. A. Starek et al. examined fuel efficient autonomous usage of the CW Equations, however, this study was conducted examining algorithms, specifically a modified version of the FMT* algorithm rather than assess the differences between forced and unforced motions.\cite{Starek2016}  Research conducted by T. Carter and M. Humi evaluated the CW Equations by adding quadratic drag, however again, this study of the CW Equations did not evaluate for the fundamental model mismatch resulting in differing fuel efficiency for forced versus unforced motion.\cite{Carter2012} M. Bando and A. Ichikawa conducted research into the CW controls problem with respect to a single input and not in comparison of model mismatch.\cite{Bando2013} Comparatively, this paper examines forced and unforced motions under specific test scenarios that are likely to be utilized during autonomous RPOD missions and maneuvers.

For the purposes of this paper, a two-satellite system is studied. Both satellites maintain a stable low Earth orbit (LEO). The uncontrolled satellites, labeled the target,  maintains a stable, circular orbit, and the target is the primary focus of most RPOD mission.\cite{Petersen2024}  The controlled satellite, labeled the chaser, pursues the target to engage in RPOD maneuvering and mission operations. These operations generally occur within $500 \mathrm{km}^2$, however motion outside that range can still generally be considered. This paper examines this two-satellite system with respect to two frames: the relative Hill’s Frame and the Earth-centered inertial frame. The Hill’s Frame fixes the target satellite at the origin with the right handed frame defined as  , where the axes represent the motion in the radial, along-track, and cross-track direction, respectively.\cite{Petersen2024, Clohessy1960} The Earth-centered inertial frame fixes the center of the Earth at the origin, with the inertial, right-handed frame with axes  , where   and   are orthogonal axes. Figure 1 demonstrates graphically the Earth-centered inertial frame and the relative frame, and these frames are utilized to derive the equations of motion for the CW Equations and the Higher Fidelity model.

The contribution of this work is showcasing that fuel usage for natural, unforced motion is approximately comparable, if not more fuel efficient under specific conditions, than forced motions. This counter-intuitive discrepancy results from model mismatch from the CW Equations, and this paper provides a novel, exact quantification of the effects of this model mismatch. By demonstrating flexibility in usage of forced motion for RPOD maneuvers and missions, the trade space is opened to higher fuel-efficiency maneuvers and other considered maneuvers for a given mission objective.

\section{Notation}

As previously stated, the dynamical frame utilized for modeling RPOD consisted of two satellites, with both satellites maintaining a stable LEO. While derivation of the Clohessy-Wiltshire equations is certainly not a novel dynamical writing, the following derivation and notation emphasizes the assumptions made in the process of obtaining the Clohessy-Wiltshire equations. This permits further analysis of these assumptions in comparison to a higher-fidelity dynamical model to study the differences in required impulsive maneuvers for various RPOD operations. 

While maintaining this stable LEO, one satellite is uncontrolled, while the other is capable of performing impulsive maneuvers to alter its orbit. The uncontrolled satellite, termed the "target", maintains a strictly circular stable orbit. The necessity of this orbit remaining circular becomes apparent during derivation of the CW Equations, as a non-circular orbit adds substantial complexity. The target is the primary focus for the simulated RPOD missions discussed in this paper. The controlled satellite, termed the "chaser", pursues the target via impulsive maneuvers to engage in RPOD maneuvering, circumnavigation, and a variety of mission operations. Notably, RPOD occurs within a range of 500 km between the satellites, thus the target and chaser will always be simulated within this range. Any maneuvering outside this range is irrelevant for the purposes of RPOD, and as such, will not be considered within this paper.

The motions of these two satellites was examined via two frames of motion: the Hill's Frame and the Earth-Centered Inertial Frame. Given the nature of RPOD and the focus on the target satellite, a relative frame of motion is often utilized to analyze and understand the position and velocity of the chaser with respect to the target. Thusly, the chaser is placed as the origin of this frame, with the target performing maneuvers and circumnavigations around that origin. This right handed frame is defined by axes $F_R \; = \; \{ \;\hat{\imath}_r, \hat{\imath}_{\theta}, \hat{\imath}_h \; \}$, where the axes represent the motion in the radial axis $\hat{\imath}_r$, the along-track axis $\hat{\imath}_{\theta}$, and the cross-track axis $\hat{\imath}_h$. Of note the cross-track axis is the direction of the target's angular momentum, and the along-track axis is the direction of the target's velocity. For notation purposes, the point $t$ represents the center of mass of the target, while the point $c$ represents the center of mass of the chaser.  

In RPOD operations, the impulsive maneuvers performed by the chaser are typically modeled as instantaneous changes in velocity, referred to as $\Delta v$. These maneuvers are critical for adjusting the relative position and velocity of the chaser with respect to the target. By focusing on impulsive maneuvers rather than continuous thrust, the analysis remains consistent with the CW Equations, which assume small perturbations over short timescales. This simplification allows for efficient trajectory planning and maneuvering without introducing the complexities of continuous thrust modeling.\cite{Curtis2013}

\begin{figure}[H]
	\centering\includegraphics[width=4.25in]{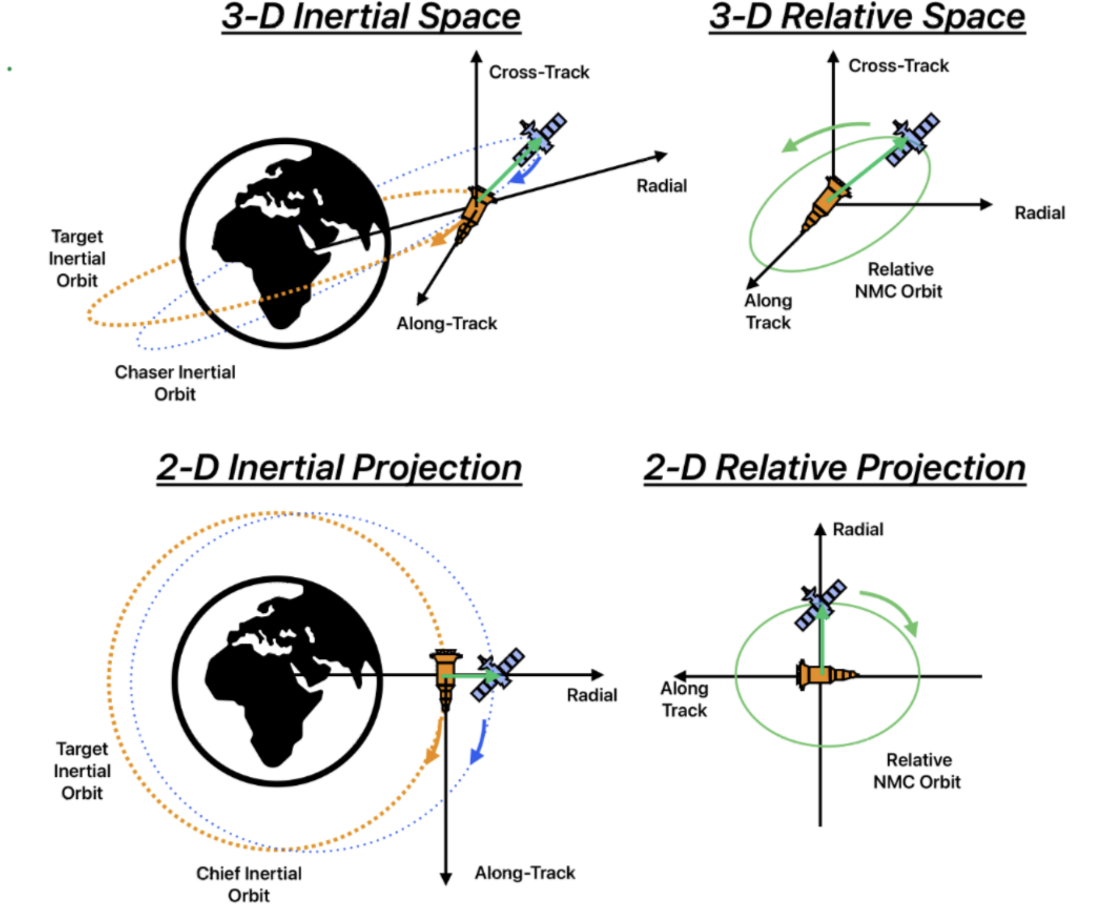}
	\caption{Graphic visualization of the Hill's Frame and the ECI Frame.}
	\label{fig:FrameVisualization}
\end{figure}

Additionally, an Earth-Centered Inertial (ECI) frame was utilized in the generation of the higher fidelity model, which made use of the two-body equations for propagation of the motion of the satellites. The gravitational forces exerted upon each satellite will be calculated and and then used to approximate the motion of both satellites. This motion was then converted into the Hill's Frame, producing a more accurate model by which to assess the motion of the chaser. The inertial frame is defined with axes $F_I \; = \; \{ \hat{\imath}, \hat{\jmath}, \hat{k} \}$, where  $\hat{\imath}, \hat{\jmath},$ and $ \hat{k}$ are the orthogonal axes governed by the right-hand rule. Again, for notation purposes, the following points are assigned. The point $e$ is located at the center of the Earth, representing the origin for the ECI Frame. The specific orientation of the X-Y plane, corresponding with the $\hat{\imath}$ and $ \hat{\jmath}$ axes, is irrelevant given that the target is assumed to be maintaining a circular orbit. However, the orientation of the $\hat{k}$ axis must align with the orientation of the $\hat{\imath}_h$ axis; the z-axes for both frames must align. These frames are depicted in Figure \ref{fig:FrameVisualization}.

\section{Clohessy-Wiltshire Equations}

Prior to derivation, several preliminary assumptions are in order to simplify dynamics and reduce the complexity of derivations. The first core assumptions required are as follows: 

\begin{quote}
\textbf{(A1)} Satellites and the Earth do not lose mass, with the Earth having significantly larger mass than either of the satellites.

\textbf{(A2)}  The Earth and satellites are spherical with equal mass distribution, thus, all three objects can be approximated as point masses.

\textbf{(A3)}  Perturbation forces on the target are negligible, and the only relevant external force is Earth’s gravity.

\textbf{(A4)}  Two restricted two-body problems are established, where the gravitational force from the satellites on the Earth and the satellites on each other is assumed to be negligible. 
\end{quote}

Assumption \textbf{(A1)} allows simplifications by ignoring mass loss, which is reasonable particularly for missions of short duration and when satellites use minimal quantities of fuel. Assumption \textbf{(A2)} enables the satellites to be treated as point masses, which is reasonable given the substantial size of the Earth in comparison to the satellites. Assumption \textbf{(A3)} simplifies the consideration of the forces exerted on the satellites to only consider the gravitational force exerted by the Earth. Together, Assumptions \textbf{(A1)} and \textbf{(A3)} were used to reasonably assume \textbf{(A4)}. Given that the Earth is substantially larger than the satellites, as noted in \textbf{(A1)}, the resultant acceleration from the imposed gravitational force would be negligible. This is also required to assume the Earth can function as an inertial frame. Further, given that the gravitational force is determined by the product of the masses of both objects, the gravitational force between the two satellites is effectively negligible in comparison to the gravitational force between the Earth and the individual satellites, since the Earth is substantially larger as assumed by \textbf{(A1)}.

Further, Assumption \textbf{(A2)} simplifies calculations by rendering the Earth as perfectly spherical. This means that the magnitude of the gravitational force for any point in three-dimensional space can be calculated simply using the orbital radius. Since the Earth is perfectly spherical with equal mass distribution, the gravitational force is only dependent on the magnitude of the orbital radius.

Under the classical dynamical model in an inertial frame, the position of both the target and the chaser can be modeled as follows, where Equations (\ref{eq:Target2Body}) and (\ref{eq:Chaser2Body}) represent the acceleration vectors of the target and chaser respectively, where $t \geq 0$ is time in seconds, $r_t(t) \in \mathbb{R}^3$, $r_c(t) \in \mathbb{R}^3$, and $U_C(t) \in \mathbb{R}^3$.

\begin{equation}
    \ddot{r}_{t/e}\; (t) \; +\; \frac{\mu}{\lVert r_t^3(t) \rVert}\; r_t(t) = 0
    \label{eq:Target2Body}
\end{equation}

\begin{equation}
    \ddot{r}_{c/e}\; (t) \; +\; \frac{\mu}{\lVert r_t^3(t) \rVert}\; r_t(t) \; - \; \frac{U_C (t)}{m_c} = 0
    \label{eq:Chaser2Body}
\end{equation}

Equations (\ref{eq:Target2Body}) and (\ref{eq:Chaser2Body}) express acceleration vectors in the ECI frame with basis vectors $\{ \hat{\imath}, \hat{\jmath}, \hat{k} \}$. These basis vectors can in turn be converted into the Hill's Frame with basis vectors $\{ \;\hat{\imath}_r, \hat{\imath}_{\theta}, \hat{\imath}_h \; \}$. These basis vectors are calculated using the relative orbital radial vector and the velocity vector as shown in Equation (\ref{eq:VectorCalculation}).

\begin{equation}
    \hat{\imath}_r \;=\; \frac{r_t(t)}{\lVert r_t(t) \rVert} \quad, \quad \hat{\imath}_h \;=\; \frac{r_t \times \dot{r}_t}{\lVert r_t \times \dot{r}_t \rVert} \quad,\quad \hat{\imath}_\theta \;=\; \hat{\imath}_h \times \hat{\imath}_r
    \label{eq:VectorCalculation}
\end{equation}

The position, velocity, and acceleration vectors relative to the target were derived utilizing the ECI frame as follows in Equation (\ref{eq:Derivation1}), where all functions are linear time dependent.

\begin{equation} \label{eq:Derivation1}
\begin{aligned}
    r_r \;&=\; r_{c/e} - r_{e/t} \quad,\quad v_r \;=\; \dot{r}_r \;-\; \omega_{I/R} \times r_r \\
    a_r \;&=\; \ddot{r}_r \;-\; \dot{\omega}_{I/R} \times r_r \;-\; \omega_{I/R}\times(\omega_{I/R}\times r_r) \;-\; 2\omega_{I/R}\times v_r
\end{aligned}
\end{equation}

For notation purposes, the following variables of x, y, and z were utilized to represent the components of the position, velocity, and acceleration of the three axes for the Hill's frame, shown in Equation (\ref{eq:CWComponents}).

\begin{equation} \label{eq:CWComponents}
    \begin{aligned}
        r_r \;&=\; x\hat{\imath}_r \;+\; y\hat{\imath}_\theta \;+\; z\hat{\imath}_h \\
        v_r \;&=\; \dot{x}\hat{\imath}_r \;+\; \dot{y}\hat{\imath}_\theta \;+\; \dot{z}\hat{\imath}_h \\
        a_r \;&=\; \ddot{x}\hat{\imath}_r \;+\; \ddot{y}\hat{\imath}_\theta \;+\; \ddot{z}\hat{\imath}_h \\
    \end{aligned}
\end{equation}

In order to continue the calculations and derivations, additional assumptions are required. 

\begin{quote} 
\textbf{(A5)}  The relative distance between the target and chaser is significantly smaller than the distance from the target to the Earth, that is: $\lVert r_r \rVert \gg \lVert r_{c/e}\rVert$.

\textbf{(A6)}  The target maintains a circular orbit about the Earth, where the magnitude of the radius is constant, that is $\lVert r_{t/e} \rVert$ is constant. 

\textbf{(A7)}  The target and chaser are in the same orbital plane about the Earth.
\end{quote}

Using the Assumption \textbf{(A5)}, the second derivative of the relative orbital position can be expressed as shown in Equation (\ref{eq:Derivation2}).

\begin{equation} \label{eq:Derivation2}
    \ddot{r}_r \; \approx \; - \frac{\mu}{\lVert r_t^3 (t) \rVert} \; ( -2x\hat{\imath}_r \;+\; y\hat{\imath}_\theta \;+\; z \hat{\imath}_h )
\end{equation}

Using the Assumptions \textbf{(A6)} and \textbf{(A7)}, the expressions for $\ddot{x}$, $\ddot{y}$, and $\ddot{z}$ can be further simplified to arrive at the linear time-invariant Clohessy-Wiltshire Equations as given below in Equation \ref{eq:CWEquations}. For the purposes of these calculations, $F_x$, $F_y$, and $F_z$ are the input control forces corresponding to the x, y, and z axes. As follows, $m_c$ is the mass of the chaser satellite. Further, $n$ is defined as $n = \mu \;/\; \lVert r_t^3 (t) \rVert$. 

\begin{equation} \label{eq:CWEquations}
    \begin{aligned}
        \ddot{x} \;&-\; 3n^2x \;-\;2n\dot{y} \;=\; \frac{F_x}{m_c} \\
        \ddot{y} \;&+\; 2n\dot{x} \;=\; \frac{F_y}{m_c} \\
        \ddot{z} \;&+\; n^2z\;=\;\frac{F_z}{m_c}
    \end{aligned}
\end{equation}

The motion of the chaser satellite can be controlled for in-plane targeting through the use of the Equation (\ref{eq:Targeting}). This equation is only applicable for targeting within the x-y plane, where this function makes use of the mean motion $n$, the commanded time span to travel from the initial to the targeted points $t_s$, as well as the initial and final points, represented with $[x_0, y_0]'$ and $[x_f, y_f]'$ respectively. In order to solve for the required impulse velocity to achieve a targeted transfer, Equation (\ref{eq:Targeting}) must be rearranged with respect to the impulse velocity $[\dot{x}_0^+, \dot{y}_0^+]'$. Equation (\ref{eq:Targeting}) is only applicable for in-plane maneuvers within the relative frame.

\begin{equation} \label{eq:Targeting}
    \begin{bmatrix}
        x_f \\ 
        y_f
    \end{bmatrix}
     \;=\; 
     \begin{bmatrix}
         4-3\cos(n*t_s) & 0 \\
         6\sin(n*t_s) - 6n*t_s & 1
     \end{bmatrix}
     \begin{bmatrix}
         x_0 \\
         y_0
     \end{bmatrix}
     \;+\;
     \begin{bmatrix}
         \frac{\sin(n*t_s)}{n} & \frac{2\cos(n*t_s)-2}{n} \\
         \frac{2-2\cos(n*t_s)}{n} & \frac{4\sin(n*t_s)-3n*t_s}{n}
     \end{bmatrix}
     \begin{bmatrix}
         \dot{x}_0^+ \\
         \dot{y}_0^+
     \end{bmatrix}
\end{equation}

Additionally, the initial velocity required to obtain a stable, NMC within the CW Model is given as $\dot{y} \;=\; -2\:n\:x_0$, where $\dot{y}$ is the initial velocity for the $\hat{\imath}_\theta$ axis, $x_0$ is the initial position in the $\hat{\imath}_r$ axis, and $n$ is the mean motion of the target.\cite{Frey2018} This stable NMT produces a closed, stable orbit in the form of a 2:1 ellipse, with the semi-major axis of the ellipse aligned with the $\hat{\imath}_\theta$ axis. 

Given Assumptions \textbf{(A1)}-\textbf{(A7)}, model mismatch is introduced to this dynamical model. Existing research discusses some means to assess and correct model mismatch. Mismatch functions, for example, can determine the validity of low order system models in comparison to their higher order counterparts. This mismatch function only assesses when differences between models becomes too great and thus invalidates the calculations and predictions of the low order model.\cite{Anderson1993} There also exist methods of adjusting models in response to excessive model mismatch and parameter deviation by adjusting a hybrid direct-indirect adaptive controller using the estimated quantification of model mismatch.\cite{Joshi2009}. This paper provides a novel representation of the model mismatch under the CW Equations by quantifying the extent of model mismatch between forced and unforced motions to assess when forced motions under RPOD provide more fuel-efficient impulsive maneuvers than the theoretically unforced motions. 

\section{Results}

The computational models described above were used to evaluate the fuel requirements for forced and unforced motion during two primary RPOD maneuvers: circumnavigation of a target satellite and interception from an initial offset to a designated rendezvous point. Both maneuvers were simulated using the Clohessy-Wiltshire (CW) equations for unforced motion and a two-body orbital propagator for forced motion. The primary objective of this comparison was to assess how model mismatch affects trajectory corrections and fuel efficiency, particularly in scenarios requiring precise station-keeping or docking. Each maneuver was analyzed across multiple maneuver sizes and numbers of impulses to determine trends in fuel efficiency and to identify cases where forced motion provided advantages over unforced trajectories. Given the importance of fuel conservation in RPOD operations, these comparisons offer insight into optimal trajectory planning for long-duration space missions.

For the circumnavigation case, the natural motion circumnavigation (NMC) trajectory followed the CW-predicted two-to-one elliptical path, defined by the initial velocity condition $\dot{y} = -2nx_0$, where $x_0$ is the initial radial offset and $n$ is the mean motion of the target satellite. The CW model assumes that this trajectory remains stable without further corrections, making it theoretically “zero-fuel.” However, deviations from the idealized CW Equations accumulate over time, particularly for larger orbits, necessitating impulse corrections to maintain the intended path. By contrast, forced motion follows a controlled circular trajectory, where impulses are applied at predefined intervals to actively maintain the orbit. While forced motion inherently consumes fuel, it also mitigates deviations caused by model mismatch, making it a more robust option when operating in regimes where CW assumptions begin to break down. The results of this study quantify how increasing trajectory size and impulse count impact fuel efficiency for both maneuver types.

The findings indicate that forced motion is generally more fuel-efficient than unforced motion, largely due to inaccuracies in the CW Equations. However, for smaller maneuver sizes, unforced motion can approach or even surpass the efficiency of forced motion when the number of impulses is increased. This suggests that frequent corrective impulses in an unforced trajectory can counteract model mismatch, improving efficiency in some cases. However, as maneuver sizes increase beyond 500 km, the discrepancy between CW-predicted and actual motion becomes more pronounced, requiring additional course corrections that reduce the efficiency of unforced trajectories. In contrast, forced motion, which consistently applies planned corrections, does not experience this same efficiency loss. These results suggest that while unforced motion may be viable for short-range proximity operations, it becomes increasingly impractical for larger orbital maneuvers where deviations from the CW model significantly impact station-keeping performance.

\begin{figure}[H]
	\centering\includegraphics[width=3.75in]{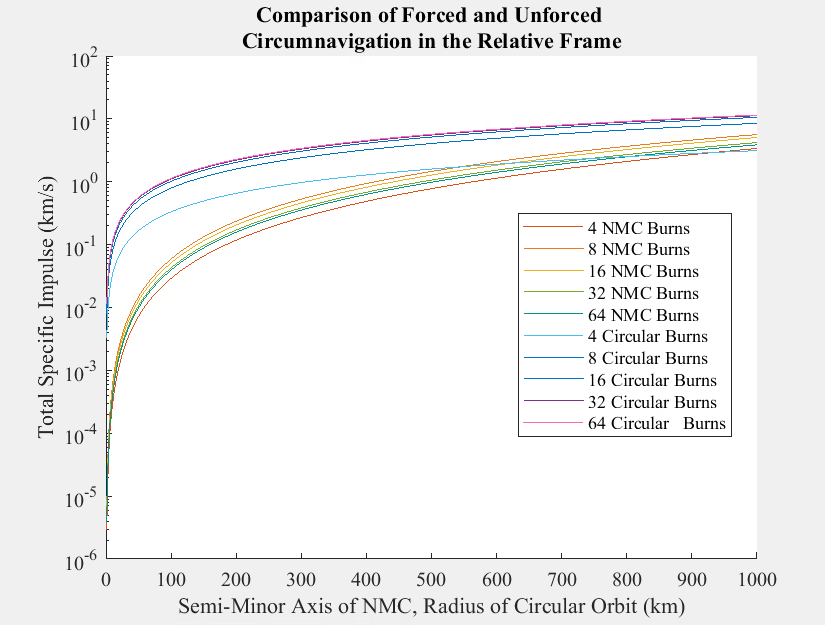}
	\caption{Comparison of required specific impulses to accomplish forced circular motion and CW NMCs with a semi-minor axis between 1 km and 1000 km.}
	\label{fig:ComparisonOverall}
\end{figure}

The total $\Delta v$ required for unforced circumnavigation using CW-based NMCs was quantified for 4, 8, 16, 32, and 64 impulsive burns, as shown in Figure \ref{fig:ComparisonOverall}. Given the order-of-magnitude differences in required impulses, a logarithmic scale was used.

\begin{figure}[H]
	\centering\includegraphics[width=3.75in]{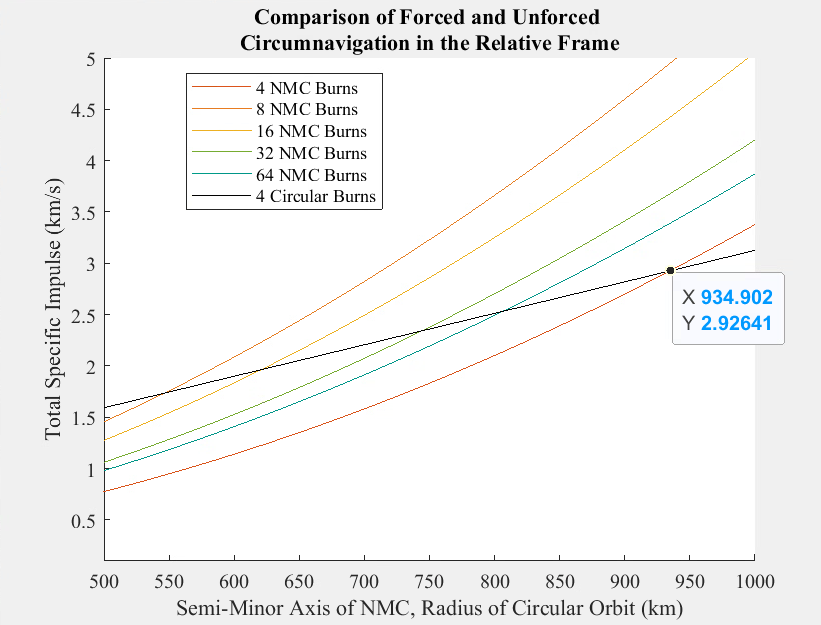}
	\caption{Quantification of required specific impulses to accomplish forced circular motion and CW NMCs with a semi-minor axis between 500 km and 1000 km.}
	\label{fig:ComparisonSpecificView}
\end{figure}

Of particular interest for the purposes of this comparison were the required impulses for the larger orbits, greater than 500 km. This plot of interest is depicted in Figure (\ref{fig:ComparisonSpecificView}. Given that these values are on approximately the same order of magnitude, a linear scale was utilized. 

At maneuver sizes greater than 500 km, the increased deviation of the CW-predicted trajectory from the actual chaser motion led to excessive corrections, ultimately making unforced motion less fuel-efficient than forced motion.

\begin{figure}[H]
	\centering\includegraphics[width=4.25in]{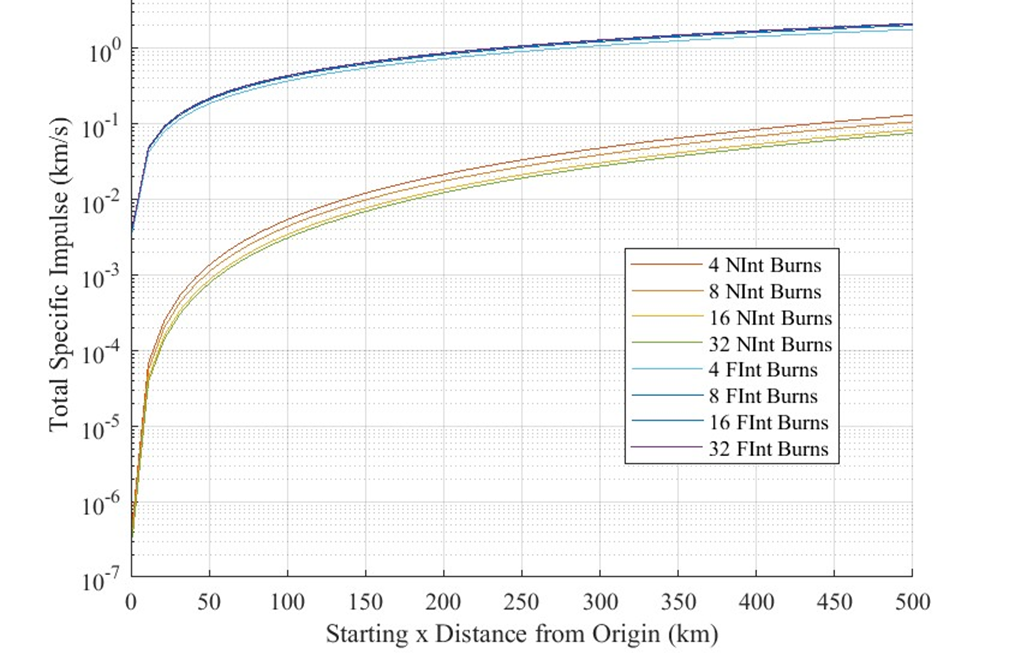}
	\caption{Comparison of required impulses for forced and unforced interception maneuvers.}
	\label{fig:InterceptImpulses}
\end{figure}

In addition to circumnavigation, a second analysis was conducted for intercept maneuvers, in which a chaser satellite performed an approach toward a designated rendezvous point. For this study, the chief satellite was placed in a circular orbit at an altitude of 2000 km, and the intercept was set to occur over a 60-minute period. Two different approaches were compared: the unforced trajectory, in which the chaser followed a CW-predicted path, and the forced trajectory, in which the chaser followed a direct linear path requiring impulse corrections at each targeting point. Unlike the circumnavigation case, where unforced motion could, in some cases, be more fuel-efficient, the results for interception maneuvers showed that unforced motion was consistently more efficient than forced motion. The forced trajectory required greater total $\Delta v$ due to the need for frequent course corrections, whereas the unforced CW trajectory naturally conformed to the orbital motion and required fewer adjustments. However, despite the increased fuel requirements, forced interception remained a viable option due to safety considerations. A forced trajectory provides greater control over the chaser’s approach, which is particularly important in high-risk operations such as docking or debris capture. The comparative impulse requirements for interception are shown in Figure \ref{fig:InterceptImpulses}.

The impact of model mismatch on fuel efficiency was a key factor influencing the results of both maneuver types. When the assumptions underlying the CW Equations began to break down, particularly at larger relative distances, unforced motion became increasingly inefficient due to accumulating deviations from the predicted trajectory. At small scales, however, the CW model retained enough accuracy that unforced motion could be a viable alternative, particularly when more frequent impulse corrections were introduced. The number of impulses played a significant role in these trends. In forced motion, increasing the number of impulses did not provide substantial benefits in fuel efficiency. However, in unforced motion, increasing the number of impulses led to improved efficiency, likely because smaller corrections minimized the effects of model mismatch over time.

It is important to note that these results were obtained by comparing the CW Equations with a two-body problem propagator. While the findings suggest that forced motion is generally more fuel-efficient due to the limitations of the CW model, this trend may not hold when additional levels of fidelity are introduced. Higher-order dynamical models, which incorporate effects such as gravitational perturbations, atmospheric drag, or third-body influences, could alter the relationship between forced and unforced efficiency. Reducing the number of simplifying assumptions and including more precise force models may reveal conditions under which forced motion becomes less efficient than previously expected. For example, the introduction of non-conservative forces or more complex gravitational effects could make forced motion more susceptible to increased fuel consumption, whereas certain corrections applied to unforced motion might allow it to maintain efficiency over longer time periods. The continued refinement of RPOD models, particularly through the incorporation of higher-fidelity numerical methods, may therefore shift the balance of efficiency between forced and unforced motion under specific circumstances.

Ultimately, while natural motion circumnavigation can sometimes be more fuel-efficient, it is not inherently superior to forced motion in all cases. The efficiency trade-offs depend on the maneuver size, the number of impulses, and the level of model mismatch. The results presented in this study suggest that further investigation using higher-fidelity models is necessary to determine whether these trends persist when additional forces and perturbations are considered. Future research should explore whether forced motion retains its advantage under more complex dynamical conditions and how real-time adaptive control methods could further optimize fuel efficiency in RPOD operations.

\section{Conclusion}

This study compared the fuel efficiency of forced and unforced motion in Rendezvous, Proximity Operations, and Docking (RPOD) maneuvers using the Clohessy-Wiltshire (CW) equations and a two-body propagator. The two primary maneuvers examined were circumnavigation, in which a chaser satellite follows a stable relative orbit around a target, and interception, in which a chaser moves from an initial offset to a rendezvous point. By quantifying the required impulse maneuvers for both forced and unforced trajectories across a range of maneuver sizes and impulse frequencies, this study provided insight into how model mismatch influences trajectory corrections and fuel efficiency. While CW-based natural motion circumnavigation (NMC) is often assumed to be fuel-efficient due to its theoretically zero-fuel stability, the results demonstrated that model mismatch significantly impacts its efficiency, particularly for larger maneuver sizes. Similarly, forced motion, while inherently requiring active corrections, proved to be more fuel-efficient in many cases due to its ability to counteract deviations introduced by the limitations of the CW model.

The key findings of this study revealed distinct trends in how forced and unforced motion scale with increasing maneuver size. Forced circumnavigation exhibited a near-linear trend in fuel consumption as maneuver size increased, whereas unforced circumnavigation tended toward an exponential increase in required impulses due to growing deviations from the CW-predicted trajectory. Although unforced motion could achieve comparable efficiency under certain conditions, particularly when the number of impulses was increased, it generally became less practical as maneuver size grew beyond 500 km. In contrast, interception maneuvers exhibited consistently worse fuel efficiency for forced motion compared to unforced motion, with forced trajectories requiring more frequent course corrections to maintain a direct approach. However, the reduced fuel efficiency of forced interception is often acceptable in operational scenarios due to the increased control and predictability it provides, which is critical for safe docking or debris capture. Unlike circumnavigation, where impulse efficiency varied significantly based on trajectory size, the trends in required impulse maneuvers for interception remained consistent across different time spans, indicating a fundamental difference in how model mismatch impacts these two maneuver types.

A key takeaway from this study is that forced motion becomes more fuel-efficient when CW model assumptions begin to break down. As relative distance increases, the inaccuracies in CW-derived unforced trajectories accumulate, requiring frequent corrective maneuvers that erode their theoretical efficiency advantage. Additionally, while increasing the number of impulses improved fuel efficiency for unforced motion, it had the opposite effect for forced motion. This suggests that while fine-grained corrections can help unforced trajectories maintain efficiency, forced trajectories are best optimized with larger, less frequent burns. Furthermore, circumnavigation maneuvers were more affected by model mismatch than interception, as the long-duration stability required for circumnavigation was more sensitive to deviations from the ideal CW-derived paths.

Future work will focus on reducing the number of simplifying assumptions in order to compare these trajectories against higher-fidelity models. The next step in this research is to incorporate additional perturbation forces, potentially including, when applicable, atmospheric drag, third-body influences, and gravitational harmonics, to more accurately assess trajectory performance under real orbital conditions. By transitioning from the two-body problem to a more comprehensive dynamical model, it will be possible to determine whether the efficiency trends observed in this study persist or if additional corrections further shift the balance between forced and unforced motion. Additionally, future studies will expand the range of test cases to include non-circular orbits, as well as evaluate how model mismatch influences trajectory deviations over extended mission durations. A higher-fidelity assessment of RPOD maneuvers will allow for more precise trajectory optimization and improved decision-making in fuel-critical space operations.

In conclusion, this study provides a systematic comparison of forced and unforced RPOD trajectories, demonstrating that while CW-based unforced motion can be fuel-efficient under specific conditions, forced motion generally remains the more reliable option when model mismatch is significant. The insights gained from this research contribute to a better understanding of how model assumptions impact trajectory planning and fuel optimization, offering valuable guidance for future RPOD mission designs.

\bibliographystyle{AAS_publication}   
\bibliography{references}   

\end{document}